\documentclass[aip,apl, preprint, numerical]{revtex4-1}

\usepackage[version=3]{mhchem} 
\usepackage[T1]{fontenc}       
\usepackage{graphicx}
\usepackage{epstopdf}
\usepackage{dcolumn}
\usepackage{bm}
\usepackage{color}
\usepackage{ulem}

\begin{document}

\title{Dielectric nanoantenna as an efficient and ultracompact demultiplexer for surface waves}

\author{Ivan~S.~Sinev}
\affiliation{Department of Nanophotonics and Metamaterials, ITMO University 197101 St. Petersburg, Russia}
\email{i.sinev@metalab.ifmo.ru}
\author{Andrey~A.~Bogdanov}
\affiliation{Department of Nanophotonics and Metamaterials, ITMO University 197101 St. Petersburg, Russia}
\author{Filipp~E.~Komissarenko}
\affiliation{Department of Nanophotonics and Metamaterials, ITMO University 197101 St. Petersburg, Russia}
\affiliation{St. Petersburg Academic University, 194021 St. Petersburg, Russia}
\author{Kristina~S.~Frizyuk}
\affiliation{Department of Nanophotonics and Metamaterials, ITMO University 197101 St. Petersburg, Russia}
\author{Mihail~I.~Petrov}
\affiliation{Department of Nanophotonics and Metamaterials, ITMO University 197101 St. Petersburg, Russia}
\author{Ivan~S.~Mukhin}
\affiliation{Department of Nanophotonics and Metamaterials, ITMO University 197101 St. Petersburg, Russia}
\affiliation{St. Petersburg Academic University, 194021 St. Petersburg, Russia}
\author{Sergey~V.~Makarov}
\affiliation{Department of Nanophotonics and Metamaterials, ITMO University 197101 St. Petersburg, Russia}
\author{Anton~K.~Samusev}
\affiliation{Department of Nanophotonics and Metamaterials, ITMO University 197101 St. Petersburg, Russia}
\author{Andrei~V.~Lavrinenko}
\affiliation{Department of Nanophotonics and Metamaterials, ITMO University 197101 St. Petersburg, Russia}
\affiliation{Department of Photonics Engineering, Technical University of Denmark, 2800 Kongens Lyngby, Denmark}
\author{Ivan~V.~Iorsh}
\affiliation{Department of Nanophotonics and Metamaterials, ITMO University 197101 St. Petersburg, Russia}


\begin{abstract}
Nanoantennas for highly efficient excitation and manipulation of surface waves at nanoscale are key elements of compact photonic circuits. However, previously implemented designs employ plasmonic nanoantennas with high Ohmic losses, relatively low spectral resolution, and complicated lithographically made architectures. Here we propose an ultracompact and simple dielectric nanoantenna (silicon nanosphere) allowing for both directional launching of surface plasmon polaritons on a thin gold film and their demultiplexing with a high spectral resolution. We show experimentally that mutual interference of magnetic and electric dipole moments supported by the dielectric nanoantenna results in opposite propagation of the excited surface waves whose wavelengths differ by less than 50~nm in the optical range. Broadband reconfigurability of the nanoantennas operational range is achieved simply by varying the diameter of the silicon sphere. Moreover, despite subwavelength size ($<\lambda/3$) of the proposed nanoantennas, they demonstrate highly efficient and directional launching of surface waves both in the forward and backward directions with the measured front-to-back ratio having a contrast of almost two orders of magnitude within a 50~nm spectral band. Our lithography-free design has great potential as highly efficient, low-cost, and ultracompact demultiplexer for advanced photonic circuits.
\end{abstract}

\maketitle

Metal surfaces\cite{raether1988SPP,kitson1996full}, metasurfaces\cite{murray2007plasmonic,smolyaninov2007magnifying,high2015visible,yermakov2015hybrid}, and 2D materials~\cite{grigorenko2012graphene,xia2014two,basov2016polaritons} provide means to support surface waves with tailorable polarization states, which fit perfectly into the emerging field of on-chip optical communications. One of the key elements for devices operating in 2D is an efficient and compact source of surface waves. On the other hand, a demultiplexer of the surface waves is another crucial element, providing parallel operation at multiple wavelengths and accelerating performance of advanced photonic circuits. 
Usually, optically driven launching of surface waves or their demultiplexing is performed using gratings~\cite{raether1988SPP}, structured nanoslits\cite{lopez2007efficient,rodriguez2013near,Yao2015}, arrays of nanoholes~\cite{drezet2007plasmonic, van2009nanohole, li2011broad,lin2013polarization} or other linear defects. However, severely limited amount of space available on a modern integrated optical circuit calls for using more compact structures for surface waves excitation and routing. 

Subwavelength particles acting as conduits between free propagating waves and waveguide modes~\cite{guo2015plasmonic} or surface waves\cite{sondergaard2004surface,evlyukhin2005point,evlyukhin2007surface,mueller2013asymmetric}, in particular, surface plasmon polaritons (SPP), have been drawing significant attention during the last decade. Coupling of far-field radiation to surface waves in this case is mediated via the near-field of induced dipole moments. Interestingly, while emission of a single dipole in 3D is inherently symmetric, its emission into surface modes can produce highly asymmetric radiation patterns. In particular, interference of two electric dipole modes with different orientations with respect to the metal surface (``rotating dipole'' configuration) was shown to provide directional launching of SPPs~\cite{rodriguez2013near,mueller2013asymmetric,xi2014unidirectional} which was the direct consequence of the photonic spin Hall effect inherent to surface waves~\cite{Bliokh2015}. Similar performance was achieved using magnetic dipole response produced by paired magnetic nanoantennas\cite{liu2012compact} and structured plasmonic couplers\cite{Pors2014,yao2016controlling}. However, fabrication of such complicated designs is a major limiting factor for their implementation.

\begin{figure}
\includegraphics[width=0.4\linewidth]{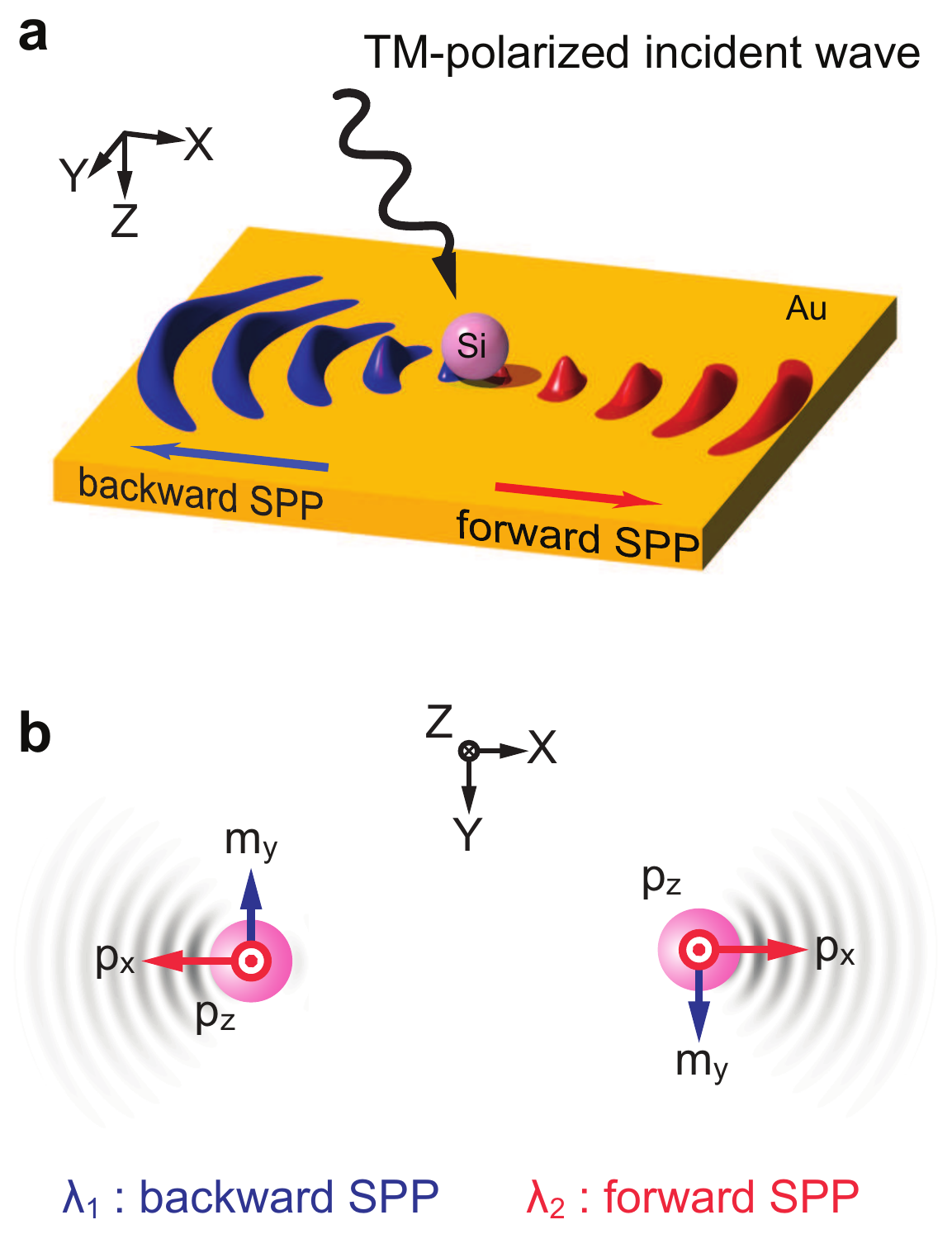}
\caption{\label{fig:scheme} (a) Scheme of SPP generation by a silicon nanosphere excited with p-polarized light at oblique incidence. (b) Scheme illustrating the spectral switching of directional launching of SPP due to different regimes of interference between dipole moments induced in the nanosphere.}
\end{figure}

A more natural way to achieve a magnetic dipole response in optics is to employ low-loss high-index dielectric nanoparticles \cite{Kuznetsov2012,evlyukhin2012demonstration,kuznetsov2016science}. Due to strong displacement currents induced in the volume of the particle, its effective magnetic polarizability can reach amplitudes comparable or even stronger than that of electric polarizability.\cite{fenollosa2008,XifrePerez2011,Kuznetsov2012} Furthermore, the spectral position and relative strength of electric and magnetic optical resonances can also be independently  tailored by changing the shape of the nanoparticle, which can lead to a plethora of fascinating effects, like directional scattering\cite{geffrin2012ncomm,fu2013ncomm,person2013demonstration,yu2015high,Decker2015}, generalized Brewster effect\cite{paniagua2016generalized}, \textit{etc.}\cite{kuznetsov2016science} However, up to now the studies of high-index nanoparticles have been mainly focused on their applications for manipulation of bulk waves, while their performance for launching surface waves remained mostly unexplored, with only a few papers addressing this problem theoretically\cite{Evlyukhin2006JETP, evlyukhin2015resonant,li2017all}.

Here, we experimentally demonstrate that a very basic dielectric nanoantenna, a single silicon nanosphere, can serve as a highly efficient and ultracompact source and demultiplexer for surface waves with a high spectral resolution ($<$50~nm) and opposite propagation directions of the demultiplexed waves. We show that the inherently strong magnetic dipole response of the silicon particle provides extremely efficient excitation of a SPP wave, whereas mutual interference of the electric and magnetic dipole resonances of the nanoantenna provides high front-to-back ratio contrast and directivity values for the excited surface waves.
Therefore, dielectric nanoantenna holds a major advantage over previously proposed SPP demultiplexer designs, offering both extreme compactness as compared to grating structures\cite{tanemura2011multiple,li2011broad,lin2013polarization} and decisively superior spectral resolution as compared to other subwavelength demultiplexer solutions.\cite{liu2011submicron, lee2014switchable,guo2015plasmonic,lu2016integrated}
The theoretical framework that we formulate to describe the physics behind the observed phenomena can be easily extended to more complex interfaces supporting other types of surface waves~\cite{high2015visible,yermakov2015hybrid}. The combination of dispersion engineering of states supported by a 2D interface and tunability of the nanoantenna radiation pattern would be an important step forward to efficient routing of surface waves.

\begin{figure}
\includegraphics[width=\linewidth]{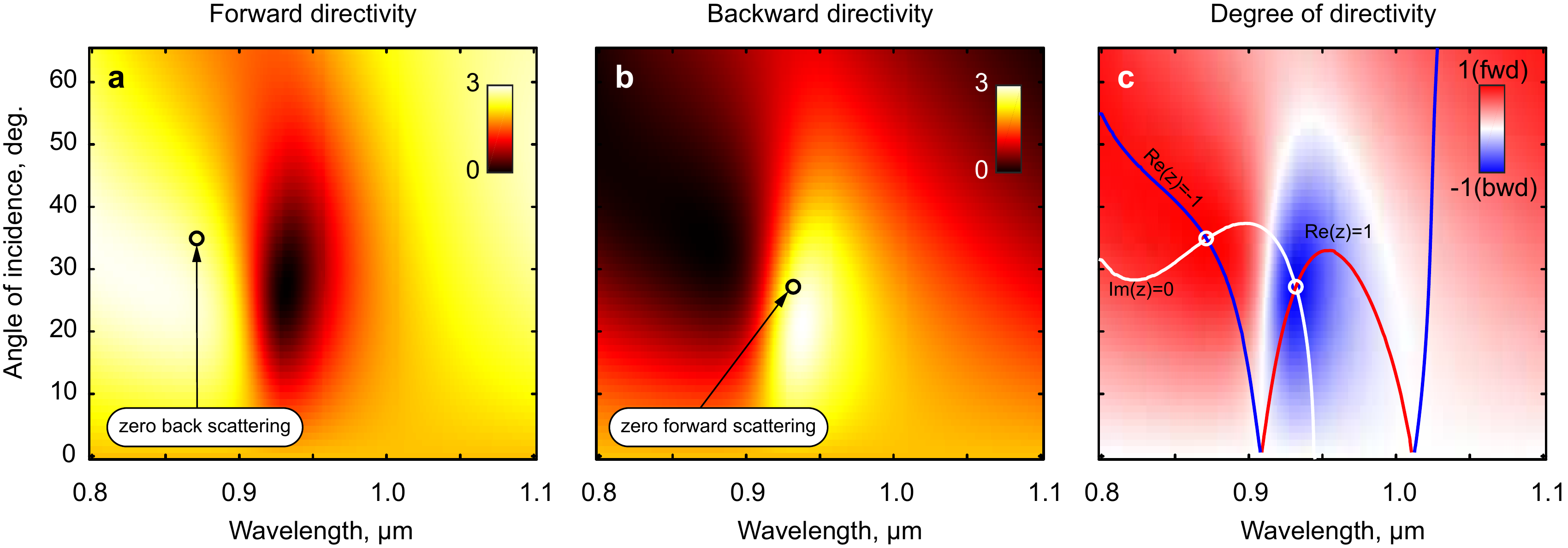}
\caption{\label{fig:analytics} Analytically calculated maps of (a) forward directivity, (b) backward directivity and (c) degree of directivity of surface plasmon polariton launched by a single 297~nm silicon nanosphere on the gold substrate. The SPP direction is given with respect to the direction of the in-plane component of the wavevector of the incident plane wave. In panel (c), the isolines for the real and imaginary parts of parameter $z$ characterizing the SPP directivity illustrate the regimes of complete cancellation of forward or backward SPP scattering at the intersections of $Re(z)=\pm1$ curves with $Im(z)=0$ curve, which are marked with circles in all three panels.}
\end{figure}

\section*{Results}

\subsection*{Theoretical concept}
To study surface plasmon polaritons excited by a silicon nanosphere, we employ the analytical model based on the Green function approach\cite{Miroshnichenko2015Bianisotropy}. This model relies on calculation of the sphere electric and magnetic polarizabilities in the dipole approximation. Applicability of the dipole approximation for calculation of plasmon fields was discussed in Ref.~\citenum{evlyukhin2005point}. Due to the structure of SPP fields,\cite{raether1988SPP} the only dipole component that does not couple to a SPP mode is the normal magnetic one. Therefore, for the s-polarized excitation the SPP directivity pattern is inherently symmetric. The p-polarized excitation, on the other hand, can provide directional excitation of a SPP due to interference between the induced dipole moments (see Fig.~\ref{fig:scheme}), as it is devised in the following.

In the case of a p-polarized wave, three independent dipole moments are excited in the silicon nanosphere: magnetic dipole along the $y$ axis, characterized by  magnetic polarizability $m_y$ and two electric dipoles along the $z$ and $x$ axes with polarizabilities $p_z$ and $p_x$, respectively (see scheme in Fig.~\ref{fig:scheme}b). Notably, polarizabilities $m_y,p_z,p_x$ differ substantially from those of a silicon nanosphere in vacuum due to bianisotropy induced by the substrate as discussed in Refs.~\citenum{Miroshnichenko2015Bianisotropy,sinev2016polarization}. The magnetic field produced by these point dipoles can be written as
\begin{align}
\mathbf{H}(\mathbf{r})=k_0^2G_H(\mathbf{r})\mathbf{m}+ik_0\nabla\times G_E(\mathbf{r})\mathbf{p},
\end{align}
where $\mathbf{r}=(\rho,\phi_0,0)$ is the radius vector in the plane of the interface in cylindrical coordinates, $k_0=\omega/c$, and $G_H,G_E$ are the total magnetic and electric Green functions for the dipole above the metallic substrate. The Green functions can be separated into the bare and reflected parts, the latter being solely responsible for the surface plasmon contribution. By writing the explicit form of the reflected parts of the Green functions as a two-dimensional Fourier transform and simplifying the equations using asymptotic relations (see Supplementary Material for details) we arrive at the following equation for the SPP magnetic field vector in the Cartesian basis: 
\begin{align}
\mathbf{H}_{SPP}\sim \frac{e^{ik_{SPP}\rho}}{\sqrt{\rho}}\begin{pmatrix} \frac{1}{2}\sin 2\phi_0(m_y-\tilde{k}_zp_x) +\tilde{k}_{SPP}\sin\phi_0 \\ \cos^2\phi_0(m_y-\tilde{k}_zp_x)-\tilde{k}_{SPP}\cos\phi_0 \\ 0\end{pmatrix}.\label{eq:HSPP}
\end{align}
Here, $\tilde{k}_{SPP}=k_{SPP}/k_0=\sqrt{\varepsilon_m/(\varepsilon_m+1)}$, $\tilde{k}_z=\sqrt{1-\tilde{k}_{SPP}^2}=\sqrt{1/(\varepsilon_m+1)}$, and $\varepsilon_m$ is the dielectric permittivity of the metal. The intensity of SPP launched by the nanosphere in a given direction is proportional to $|H_{\phi}|^2$, where $H_{\phi}=H_y\cos\phi_0-H_x\sin\phi_0$. We thus obtain the expression for the SPP field intensity in a given direction:
\begin{align}
\label{eq:Ispp}
I_{SPP}\sim \frac{1}{\rho} \left|\cos\phi_0 (m_y-i\kappa p_x) -\tilde{k}_{SPP}p_z\right|^2,
\end{align}

where $\kappa=-i\tilde{k}_z$. It is convenient to introduce the complex dimensionless parameter $z=(m_y-i\kappa p_x)/(\tilde{k}_{SPP}p_z)$. Then, the azimuthal dependence of the plasmon intensity can be written as:
\begin{align}
\label{eq:Isppz}
I_{SPP}(\phi_0)\sim (1-\mathrm{Re}z\cos\phi_0)^2+(\mathrm{Im}z)^2\cos^2\phi_0.
\end{align}
This expression has two important consequences. First, it can be shown that the maximal achievable directivity 
$D(\phi_0)=2\pi I_{SPP}(\phi_0)/\int d\phi I_{SPP}(\phi)$ is equal to $3$. Second, unlike the case of non-active nanoantennas for bulk waves,\cite{kerker1983electromagnetic,garcia2011directionality,geffrin2012ncomm, fu2013ncomm} both forward or backward scattering of the incident light into SPPs can be totally suppressed, if condition $z=\pm1$ is satisfied. 

Remarkably, these fascinating effects are readily available for the simplest example of a dielectric nanoantenna: nanoparticle of a spherical shape. This is illustrated in Fig.~\ref{fig:analytics} for a 297~nm silicon nanosphere on gold substrate. The maps of forward and backward directivity  shown in Fig.~\ref{fig:analytics}a,b in the ``wavelength-angle of incidence'' axes reveal resonant suppression of forward SPP and respective increase of  the backward SPP at around 935~nm. Importantly, for the presented spectral range and angles of incidence the conditions for total suppression of backward and forward scattering are independently fullfilled. This is best illustrated in Fig.~\ref{fig:analytics}c, where the map of the degree of directivity (DOD) is presented. We define this value as $\text{DOD}=(I_f-I_b)/(I_f+I_b)$, where $I_f$ and $I_b$ stand for the SPP intensity in the forward and backward directions, respectively. Following Eq.\eqref{eq:Isppz}, the exact excitation conditions for total cancellation of backward (DOD=1) and forward (DOD=-1) SPPs can be found as the intersections of $Im(z)=0$ isoline with either $Re(z)=-1$ or $Re(z)=1$ isolines, which are also shown in Fig.~\ref{fig:analytics}c. The relevant intersection points are marked with circles in Fig.~\ref{fig:analytics}a-c.

\subsection*{Experimental results}

\begin{figure*}
\includegraphics[width=0.7\linewidth]{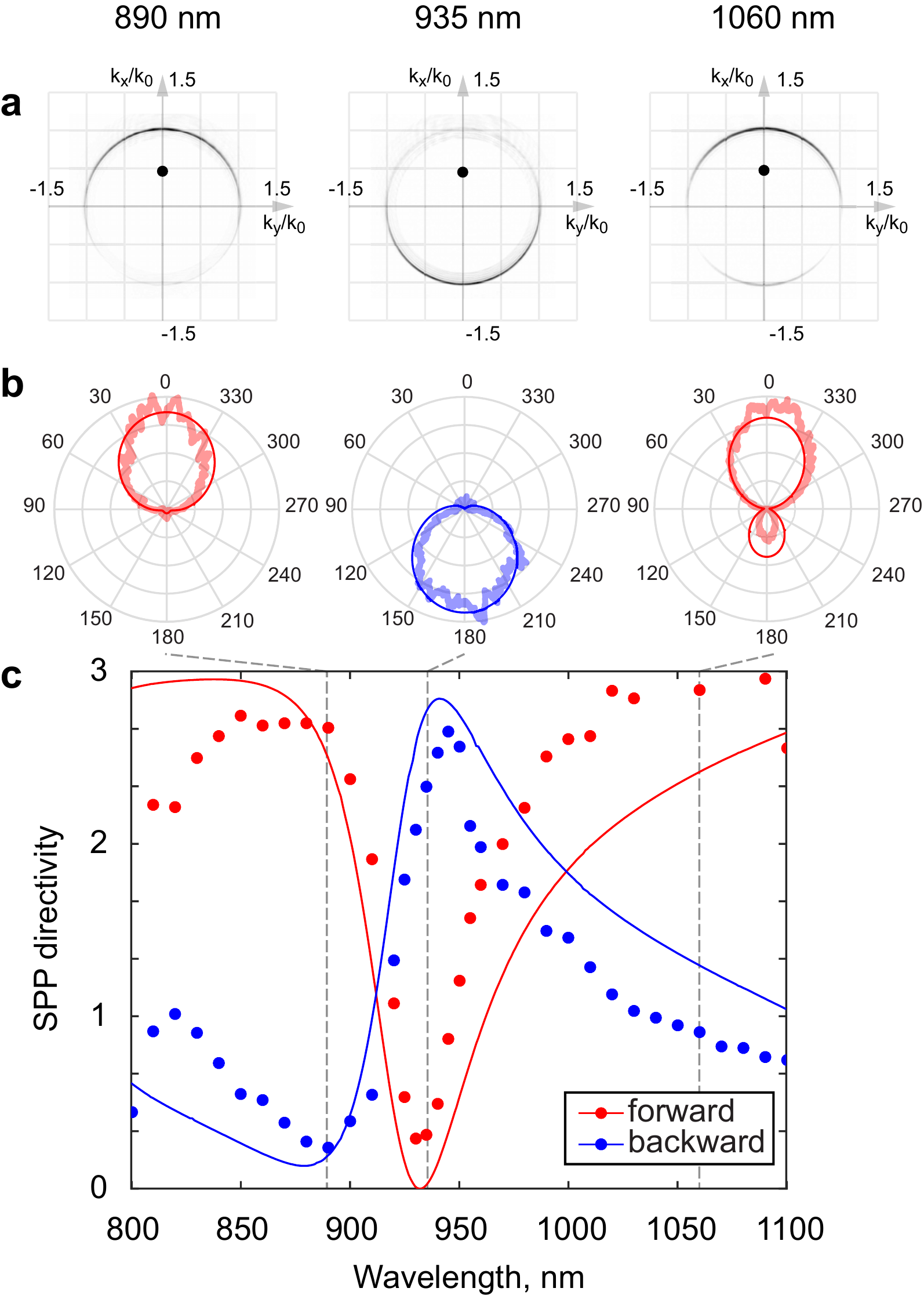}
\caption{\label{fig:experiment} Experimental demonstration of switching of SPP radiation patterns by a silicon nanoantenna. (a) (False color) Fourier plane images of SPP launched by a $\approx$295~nm silicon nanosphere on 40~nm gold film for 890, 935 and 1060~nm excitation wavelengths. Small black circles mark the angular range of the excitation radiation which is incident at around 25~degrees. (b) SPP directivity patterns reconstructed from the measured Fourier images (thick lines) and analytically calculated directivity patterns (thin lines). (c) Spectral dependence of the forward and backward SPP intensity demonstrating efficient switching between the SPP excitation directions. Experimental data (dots) and analytical data for a 297~nm silicon sphere on the gold substrate are shown. The wavelengths corresponding to data presented in (a,b) are marked with dashed lines.}

\end{figure*}
 
To confirm the SPP directivity switching experimentally, we visualized the directivity patterns of SPPs excited on a thin gold film by a single silicon nanosphere using a leakage radiation microscopy (LRM) setup combined with Fourier plane imaging optics\cite{hecht1996local, drezet2008leakage, zhang2010direct, frisbie2010characterization} (for details see Methods and Supplementary Material). Here, we focused on the observation of the regime of forward SPP scattering cancellation, since it is a crucial feature of our system. Therefore, the angle of incidence of a p-polarized beam was chosen to be 25~degrees, which allows the fulfillment of condition $z=1$ according to the analytical model (see Fig.~\ref{fig:analytics}c). Fig.~\ref{fig:experiment}a,b show the Fourier plane images and reconstructed directivity patterns of SPPs from a single silicon nanosphere (d$\approx$295~nm) for three distinctive regimes at characteristic excitation wavelengths: highly directional forward scattering (890~nm), inversion of the directivity pattern at 935~nm, and recovery of forward scattering at 1060~nm. The resonant behavior of the switching process is illustrated in Fig.~\ref{fig:experiment}c, where the spectral dependence of the SPP directivity in the forward and backward directions is shown. These results are in excellent agreement with the analytical calculations of the SPP directivity performed using Eq.\eqref{eq:Ispp} for a 297~nm silicon nanosphere, which are shown in Fig.~\ref{fig:experiment}c with thin lines. The sphere diameter in the calculations was chosen for the best matching of the spectral position of forward SPP directivity dip near 935~nm.  Note that unlike the analytical data, the measured minimum forward plasmon intensity  does not reach zero due to the angular divergence and non-monochromaticity of the incident beam. Despite this, we readily observe switching of front-to-back ratio of SPP excitation from 11 to 0.13 within less than 50~nm spectral band (890~ to 935~nm). The full spectral dependence of the measured Fourier plane images for the 295~nm nanosphere is shown in Supplementary Movie S1.

\section*{Discussion}

\begin{figure*}
\includegraphics[width=\linewidth]{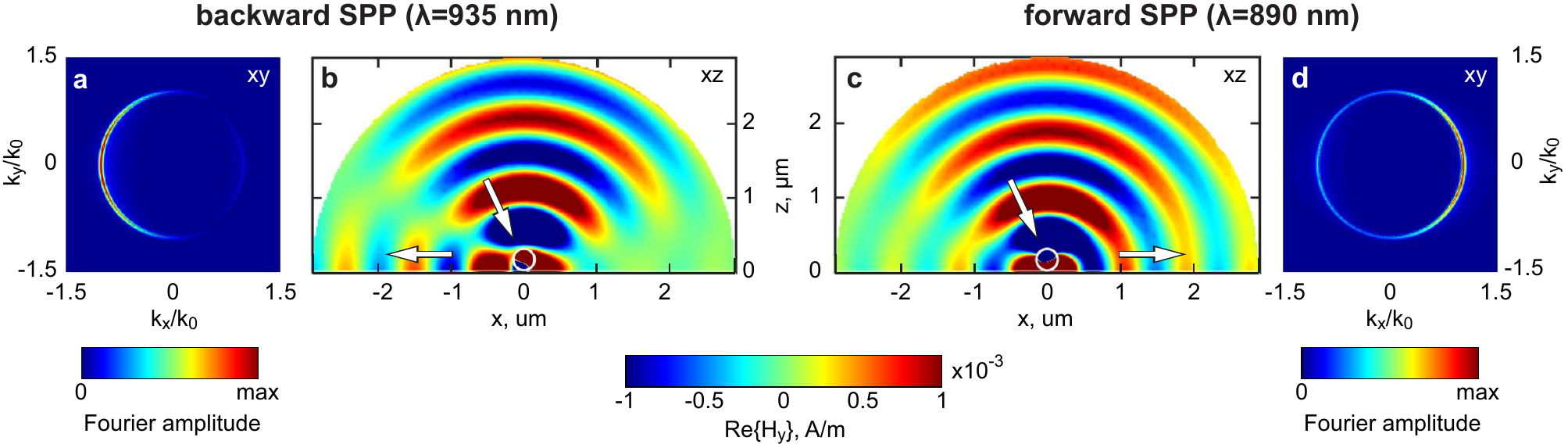}
\caption{\label{fig:COMSOL} Numerical calculations illustrating the backward (a,b), $\lambda$=935~nm and forward (c,d), $\lambda$=890~nm, SPP excitation regimes by a 290~nm silicon nanosphere on 40~nm gold film excited by a plane wave incident at 25~degrees to the substrate surface. (b) and (c) show the sections of the scattered magnetic field ($H_y$) in the plane of incidence (xz). The directions of the incident plane wave and the the SPP launched by the sphere are shown with arrows. (a) and (d) show the Fourier transforms of the calculated SPP fields in the substrate plane (xy) illustrating the directivity of the excited SPPs.}
\end{figure*}

We also illustrate the demultiplexing of SPPs and the applicability of the dipole approximation with full wave numerical calculations of the electromagnetic fields distribution near the silicon nanoantenna on the gold film. The results of the simulations for the backward ($\lambda$=935~nm) and forward ($\lambda$=890~nm) SPP excitation regimes for 25$^{\circ}$ angle of incidence are presented in Fig.~\ref{fig:COMSOL}. The maps of the scattered magnetic field distribution in the plane (Fig.~\ref{fig:COMSOL}b,c) unambiguously show the switching of the SPP direction. This is further supported by the Fourier transform of the SPP fields presented in Fig.~\ref{fig:COMSOL}a,d (see Supplementary Material), which perfectly reproduce the experimental leakage radiation images (Fig.~\ref{fig:experiment}). Note that  in numerical simulations the best matching of the SPP reversal condition to the experimental data was achieved for the sphere diameter of 290~nm, which slightly differs from the sphere diameter in analytical model (297~nm).  Time evolution of the scattered magnetic field profiles in forward and backward SPP excitation regimes are illustrated in Supplementary Movies S2 and S3, respectively.

\begin{figure*}
\includegraphics[width=0.9\linewidth]{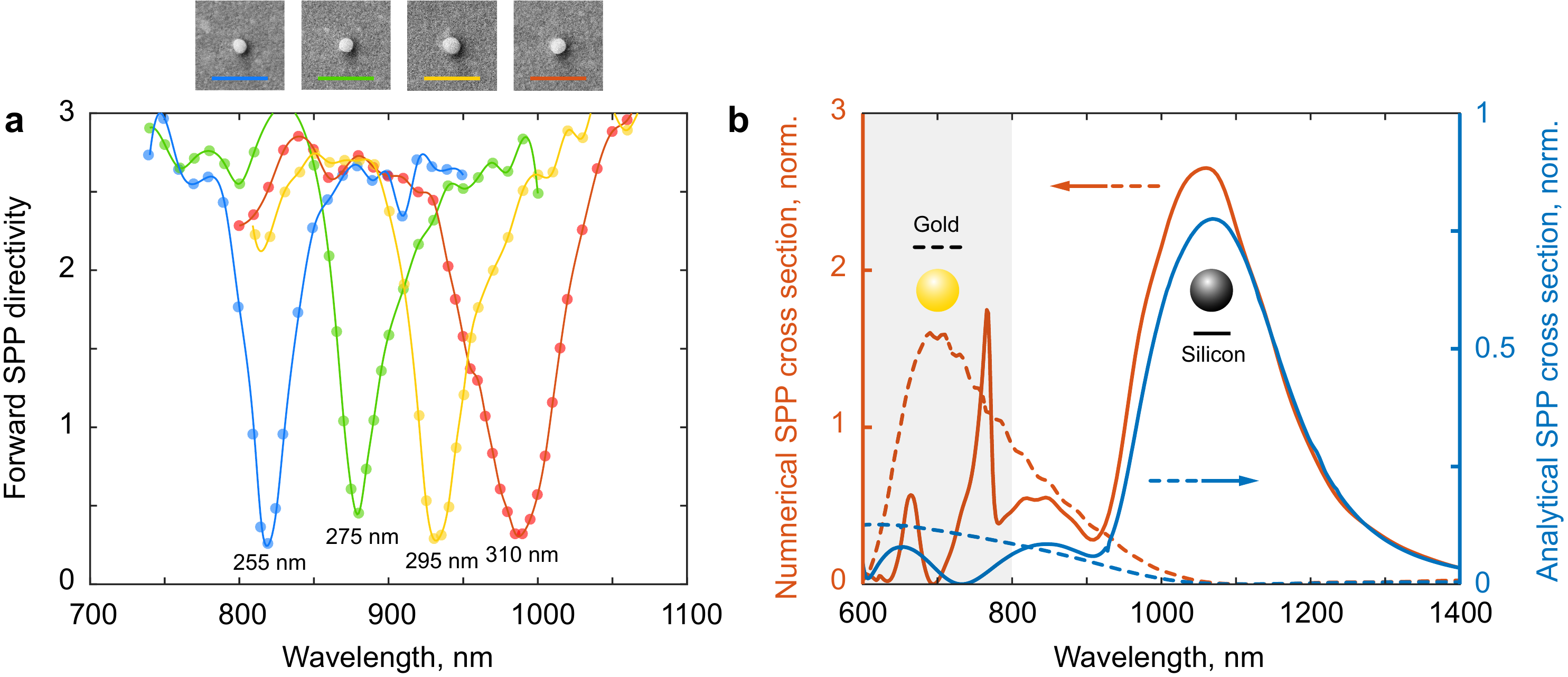}
\caption{\label{fig:scalability} 
(a) Experimentally measured spectra of forward SPP directivity for four silicon nanospheres of different diameters showing the scalability of the SPP switching condition. The inset shows SEM images of the particles in the order of increasing the wavelength of the measured forward SPP directivity dip. This is additionally encoded with the color of the scale bars which represent 1~$\mu$m. The approximate sizes of the nanoparticles marked in the plot are obtained via fitting the spectra with the analytical model (not shown).  (b)~Calculated scattering cross sections into SPP for silicon (solid lines) and gold (dashed lines) nanoparticles of the same sizes reveals superior SPP excitation efficiency by a silicon nanoparticle. The results of numerical model (red lines, sphere diameter 290~nm) and analytical calculation within dipole approximation (blue lines, sphere diameter 297~nm) are compared. The cross sections are calculated for normally incident plane wave excitation and are normalized to the physical section of the particles. The shaded area shows the spectral region of excitation of higher order multipole modes, where the analytical model is not applicable.}
\end{figure*}

Another major advantage of using dielectric nanoantennas for excitation of SPPs is the scalability of their resonances and, consequently, the spectral position of the resonant cancellation of forward scattering. This is illustrated in Fig.~\ref{fig:scalability}a, where the experimental spectra of the forward SPP directivity is shown for four silicon nanospheres of different diameters ranging from 250 to 310~nm. Therefore, simply by changing the nanoparticle size, which can be controlled with high precision in modern laser ablation techniques,\cite{zywietz2014laser} resonant inversion of the SPP excitation pattern can be tuned within a broad spectral range from visible to infrared frequencies.

Moreover, the directional excitation of SPPs with dielectric nanoparticles is extremely efficient due to their strong magnetic dipole response. The  expression for the SPP intensity (see Eq.\eqref{eq:Ispp}) shows that the contribution of the lateral electric dipole moment to SPP excitation is suppressed by a factor of 
$1/\kappa\sim \sqrt{\varepsilon_m+1}$ as compared to the contribution of lateral magnetic dipole moment. For the SPP on gold in the near-infrared spectral range, this ratio can reach values of 5-7.\cite{JohnsonChristy}
To provide quantitative estimation of the SPP excitation efficiency, we compare the cross sections of scattering of plane waves into SPPs for gold and silicon nanoparticles of the same size. While silicon nanoparticles possess an inherently strong magnetic dipole response, the gold nanoparticles gain one only due to interaction with the gold substrate via the bianisotropy mechanism\cite{Miroshnichenko2015Bianisotropy}, and it is still weak as compared to the electric dipole moment. Spectral dependences of SPP excitation efficiency for gold and silicon nanospheres are compared in Fig.~\ref{fig:scalability}b. It shows the cross sections of scattering into SPP for gold and silicon nanospheres of the same size calculated for normal incidence and normalized to the physical cross sections of the nanoparticles. The results of both numerical and analytical calculations are shown. The presented data confirm the expected superior performance of silicon nanoparticles for excitation of SPPs and reveal the spectral range of applicability of the point dipole approximation.  The maximum cross section of scattering into SPP, which is reached in the vicinity of magnetic dipole resonance, is of the same order as the physical cross-section of the nanoparticle, thus comprising a significant fraction of its optical response (see Supplementary Figure~S1 for comparison between different contributions to the total extinction cross section). Importantly, Fig.~\ref{fig:scalability}b also reveals that the analytical model based on the dipole approximation, while accurately reproducing the spectral dependence of SPP excitation efficiency, underestimates its absolute value. This inconsistency is due to the fact that in analytical calculations, we set the point dipole with modified polarizability to  the center of the sphere as it is usually done in theoretical studies.\cite{evlyukhin2015resonant, Miroshnichenko2015Bianisotropy} However, when the particle is placed on gold substrate, the fields become mainly concentrated near the metal surface\cite{sinev2016polarization}. Therefore, the effective position of the dipole moment should also be closer to the metal surface, which increases the efficiency of coupling to SPP mode. This effect is especially relevant for gold nanoparticle, when the fields are mostly concentrated in the gap between nanoparticle and substrate.

To conclude, we have revealed huge potential of high-index dielectric nanoantennas for directional excitation of surface waves and their highly efficient demultiplexing. We have experimentally demonstrated resonant switching between forward and backward excitation of surface plasmon polariton modes within a narrow spectral band (less than 50~nm) by a single silicon nanosphere. We show that mutual interference of magnetic and electric dipole moments of the nanosphere can provide complete suppression of forward or backward surface waves for different excitation conditions. By merit of scalability of optical resonances of dielectric nanoantennas, these regimes can be tuned within a broad spectral range from visible to infrared. We also show that the strong magnetic dipole response of a silicon nanosphere provides a superior surface plasmon polariton excitation efficiency, which can not be reached with metallic nanospheres. Importantly, the demultiplexing nanoantennas can be fabricated by low-cost and simple laser printing technique, which allows for fast and precise nanosphere deposition on any surface, including predesigned metasurfaces and 2D materials. Keeping in mind that the resonant switching takes place at relatively small angles of incidence of the impinging beam (25$^o$), our findings have important practical implications for on-chip optical communications and surface photonics.

\section*{Methods}

\subsection*{Sample fabrication}

Silicon nanospheres were fabricated from a thin silicon film on the glass substrate using the femtosecond laser ablation method \cite{zywietz2014laser,dmitriev2016laser}. After that, nanospheres with diameters ranging from 250 to 310~nm were transferred to a thin (40~nm) gold film on the glass substrate by nanomanipulations under an electron beam\cite{Denisyuk2014}. This method allows for precise positioning of nanospheres in clear regions of the substrate far from other defects, that ensures extremely low parasitic scattering of the SPPs.

\subsection*{Experiment}

In the leakage radiation microscopy experiments, a SPP from a single silicon nanosphere was excited with an obliquely incident TM-polarized beam mildly focused on the nanosphere with an achromatic doublet lens, which ensured good trade-off between the SPP excitation efficiency and angular divergence of the excitation beam. The SPP radiation leaking through the thin gold film was then collected from the bottom with an oil immersion objective. For tracking the spectral evolution of the SPP directivity patterns, we used supercontinuum white light source (Fianium WhiteLase SC400-6) combined with acousto-optic tunable filter (Fianium AOTF) yielding a narrowband ($\approx$4~nm) beam with the tunable central wavelength within the visible to near-infrared regions (650-1100~nm). To provide a smooth Gaussian beam profile required for accurate measurement of Fourier plane images, the beam was spatially filtered with a single-mode fiber. Since the data provided by the Fourier plane image represent the angular spectrum of the SPP leakage radiation, it offers a simple and direct way of reconstructing the full SPP directivity pattern. This is achieved by plotting the the radial dependence of the intensity of light coming from the sample at an angle corresponding to the direction of SPP leakage radiation, which in the Fourier image forms a well-defined double-crescent shape (see Fig.~\ref{fig:experiment}a).

\subsection*{Numerical calculations}

Numerical calculations were performed using COMSOL Multiphysics$^{\textregistered}$ package.

\section*{Acknowledgement}

This work was financially supported by Russian Science Foundation, grant no. 15-12-20028.

The authors thank Andrey Evlyukhin and Osamu Takayama for fruitful discussions.

\section*{Author contributions}

\subsection*{Contributions} 
I.S.S. performed the experiment and analytical calculations and wrote the manuscript. A.A.B. performed the numerical calculations of the SPP excitation efficiency.
F.E.K. performed the sample fabrication, including the transfer of silicon nanoparticles on gold substrate.
K.S.F and M.I.P performed the full-wave numerical simulations.
I.S.M. supervised the sample fabrication. 
S.V.M fabricated the silicon nanoparticles.
A.K.S. participated in analytical calculations and experimental data processing.
A.V.L. supervised the RSF project.
I.V.I. conceived the idea and developed the analytical model.
All authors contributed to discussions and proofreading of the manuscript.

\subsection*{Corresponding author}
Correspondence and requests for materials should be addressed to Ivan Sinev 
(email: \url{i.sinev@metalab.ifmo.ru}).


\begin{thebibliography}{99}
\expandafter\ifx\csname url\endcsname\relax
  \def\url#1{\texttt{#1}}\fi
\expandafter\ifx\csname urlprefix\endcsname\relax\def\urlprefix{URL }\fi
\providecommand{\bibinfo}[2]{#2}
\providecommand{\eprint}[2][]{\url{#2}}

\bibitem{raether1988SPP}
\bibinfo{author}{Raether, H.}
\newblock \emph{\bibinfo{title}{Surface plasmons on smooth surfaces}}
  (\bibinfo{publisher}{Springer}, \bibinfo{year}{1988}).

\bibitem{kitson1996full}
\bibinfo{author}{Kitson, S.}, \bibinfo{author}{Barnes, W.~L.} \&
  \bibinfo{author}{Sambles, J.}
\newblock \bibinfo{title}{Full photonic band gap for surface modes in the
  visible}.
\newblock \emph{\bibinfo{journal}{Physical Review Letters}}
  \textbf{\bibinfo{volume}{77}}, \bibinfo{pages}{2670} (\bibinfo{year}{1996}).

\bibitem{murray2007plasmonic}
\bibinfo{author}{Murray, W.~A.} \& \bibinfo{author}{Barnes, W.~L.}
\newblock \bibinfo{title}{Plasmonic materials}.
\newblock \emph{\bibinfo{journal}{Advanced Materials}}
  \textbf{\bibinfo{volume}{19}}, \bibinfo{pages}{3771--3782}
  (\bibinfo{year}{2007}).

\bibitem{smolyaninov2007magnifying}
\bibinfo{author}{Smolyaninov, I.~I.}, \bibinfo{author}{Hung, Y.-J.} \&
  \bibinfo{author}{Davis, C.~C.}
\newblock \bibinfo{title}{Magnifying superlens in the visible frequency range}.
\newblock \emph{\bibinfo{journal}{Science}} \textbf{\bibinfo{volume}{315}},
  \bibinfo{pages}{1699--1701} (\bibinfo{year}{2007}).

\bibitem{high2015visible}
\bibinfo{author}{High, A.~A.} \emph{et~al.}
\newblock \bibinfo{title}{Visible-frequency hyperbolic metasurface}.
\newblock \emph{\bibinfo{journal}{Nature}} \textbf{\bibinfo{volume}{522}},
  \bibinfo{pages}{192--196} (\bibinfo{year}{2015}).

\bibitem{yermakov2015hybrid}
\bibinfo{author}{Yermakov, O.} \emph{et~al.}
\newblock \bibinfo{title}{Hybrid waves localized at hyperbolic metasurfaces}.
\newblock \emph{\bibinfo{journal}{Physical Review B}}
  \textbf{\bibinfo{volume}{91}}, \bibinfo{pages}{235423}
  (\bibinfo{year}{2015}).

\bibitem{grigorenko2012graphene}
\bibinfo{author}{Grigorenko, A.}, \bibinfo{author}{Polini, M.} \&
  \bibinfo{author}{Novoselov, K.}
\newblock \bibinfo{title}{Graphene plasmonics}.
\newblock \emph{\bibinfo{journal}{Nature photonics}}
  \textbf{\bibinfo{volume}{6}}, \bibinfo{pages}{749--758}
  (\bibinfo{year}{2012}).

\bibitem{xia2014two}
\bibinfo{author}{Xia, F.}, \bibinfo{author}{Wang, H.}, \bibinfo{author}{Xiao,
  D.}, \bibinfo{author}{Dubey, M.} \& \bibinfo{author}{Ramasubramaniam, A.}
\newblock \bibinfo{title}{Two-dimensional material nanophotonics}.
\newblock \emph{\bibinfo{journal}{Nature Photonics}}
  \textbf{\bibinfo{volume}{8}}, \bibinfo{pages}{899--907}
  (\bibinfo{year}{2014}).

\bibitem{basov2016polaritons}
\bibinfo{author}{Basov, D.}, \bibinfo{author}{Fogler, M.} \&
  \bibinfo{author}{de~Abajo, F.~G.}
\newblock \bibinfo{title}{Polaritons in van der waals materials}.
\newblock \emph{\bibinfo{journal}{Science}} \textbf{\bibinfo{volume}{354}},
  \bibinfo{pages}{aag1992} (\bibinfo{year}{2016}).

\bibitem{lopez2007efficient}
\bibinfo{author}{L{\'o}pez-Tejeira, F.} \emph{et~al.}
\newblock \bibinfo{title}{Efficient unidirectional nanoslit couplers for
  surface plasmons}.
\newblock \emph{\bibinfo{journal}{Nature Physics}}
  \textbf{\bibinfo{volume}{3}}, \bibinfo{pages}{324--328}
  (\bibinfo{year}{2007}).

\bibitem{rodriguez2013near}
\bibinfo{author}{Rodr{\'\i}guez-Fortu{\~n}o, F.~J.} \emph{et~al.}
\newblock \bibinfo{title}{Near-field interference for the unidirectional
  excitation of electromagnetic guided modes}.
\newblock \emph{\bibinfo{journal}{Science}} \textbf{\bibinfo{volume}{340}},
  \bibinfo{pages}{328--330} (\bibinfo{year}{2013}).

\bibitem{Yao2015}
\bibinfo{author}{Yao, W.} \emph{et~al.}
\newblock \bibinfo{title}{Efficient directional excitation of surface plasmons
  by a single-element nanoantenna}.
\newblock \emph{\bibinfo{journal}{Nano Letters}} \textbf{\bibinfo{volume}{15}},
  \bibinfo{pages}{3115--3121} (\bibinfo{year}{2015}).


\bibitem{drezet2007plasmonic}
\bibinfo{author}{Drezet, A.} \emph{et~al.}
\newblock \bibinfo{title}{Plasmonic crystal demultiplexer and multiports}.
\newblock \emph{\bibinfo{journal}{Nano Letters}} \textbf{\bibinfo{volume}{7}},
  \bibinfo{pages}{1697--1700} (\bibinfo{year}{2007}).

\bibitem{van2009nanohole}
\bibinfo{author}{Van~Oosten, D.}, \bibinfo{author}{Spasenovic, M.} \&
  \bibinfo{author}{Kuipers, L.}
\newblock \bibinfo{title}{Nanohole chains for directional and localized surface
  plasmon excitation}.
\newblock \emph{\bibinfo{journal}{Nano Letters}} \textbf{\bibinfo{volume}{10}},
  \bibinfo{pages}{286--290} (\bibinfo{year}{2009}).

\bibitem{li2011broad}
\bibinfo{author}{Li, L.}, \bibinfo{author}{Li, T.}, \bibinfo{author}{Wang, S.},
  \bibinfo{author}{Zhu, S.} \& \bibinfo{author}{Zhang, X.}
\newblock \bibinfo{title}{Broad band focusing and demultiplexing of in-plane
  propagating surface plasmons}.
\newblock \emph{\bibinfo{journal}{Nano Letters}} \textbf{\bibinfo{volume}{11}},
  \bibinfo{pages}{4357--4361} (\bibinfo{year}{2011}).

\bibitem{lin2013polarization}
\bibinfo{author}{Lin, J.} \emph{et~al.}
\newblock \bibinfo{title}{Polarization-controlled tunable directional coupling
  of surface plasmon polaritons}.
\newblock \emph{\bibinfo{journal}{Science}} \textbf{\bibinfo{volume}{340}},
  \bibinfo{pages}{331--334} (\bibinfo{year}{2013}).

\bibitem{guo2015plasmonic}
\bibinfo{author}{Guo, R.} \emph{et~al.}
\newblock \bibinfo{title}{Plasmonic fano nanoantennas for on-chip separation of
  wavelength-encoded optical signals}.
\newblock \emph{\bibinfo{journal}{Nano letters}} \textbf{\bibinfo{volume}{15}},
  \bibinfo{pages}{3324--3328} (\bibinfo{year}{2015}).

\bibitem{sondergaard2004surface}
\bibinfo{author}{S{\o}ndergaard, T.} \& \bibinfo{author}{Bozhevolnyi, S.}
\newblock \bibinfo{title}{Surface plasmon polariton scattering by a small
  particle placed near a metal surface: An analytical study}.
\newblock \emph{\bibinfo{journal}{Physical Review B}}
  \textbf{\bibinfo{volume}{69}}, \bibinfo{pages}{045422}
  (\bibinfo{year}{2004}).

\bibitem{evlyukhin2005point}
\bibinfo{author}{Evlyukhin, A.} \& \bibinfo{author}{Bozhevolnyi, S.}
\newblock \bibinfo{title}{Point-dipole approximation for surface plasmon
  polariton scattering: Implications and limitations}.
\newblock \emph{\bibinfo{journal}{Physical Review B}}
  \textbf{\bibinfo{volume}{71}}, \bibinfo{pages}{134304}
  (\bibinfo{year}{2005}).

\bibitem{evlyukhin2007surface}
\bibinfo{author}{Evlyukhin, A.}, \bibinfo{author}{Brucoli, G.},
  \bibinfo{author}{Mart{\'\i}n-Moreno, L.}, \bibinfo{author}{Bozhevolnyi, S.}
  \& \bibinfo{author}{Garc{\'\i}a-Vidal, F.}
\newblock \bibinfo{title}{Surface plasmon polariton scattering by finite-size
  nanoparticles}.
\newblock \emph{\bibinfo{journal}{Physical Review B}}
  \textbf{\bibinfo{volume}{76}}, \bibinfo{pages}{075426}
  (\bibinfo{year}{2007}).

\bibitem{mueller2013asymmetric}
\bibinfo{author}{Mueller, J.~B.} \& \bibinfo{author}{Capasso, F.}
\newblock \bibinfo{title}{Asymmetric surface plasmon polariton emission by a
  dipole emitter near a metal surface}.
\newblock \emph{\bibinfo{journal}{Physical Review B}}
  \textbf{\bibinfo{volume}{88}}, \bibinfo{pages}{121410}
  (\bibinfo{year}{2013}).

\bibitem{xi2014unidirectional}
\bibinfo{author}{Xi, Z.}, \bibinfo{author}{Lu, Y.}, \bibinfo{author}{Yu, W.},
  \bibinfo{author}{Wang, P.} \& \bibinfo{author}{Ming, H.}
\newblock \bibinfo{title}{Unidirectional surface plasmon launcher: rotating
  dipole mimicked by optical antennas}.
\newblock \emph{\bibinfo{journal}{Journal of Optics}}
  \textbf{\bibinfo{volume}{16}}, \bibinfo{pages}{105002}
  (\bibinfo{year}{2014}).

\bibitem{Bliokh2015}
\bibinfo{author}{Bliokh, K.~Y.}, \bibinfo{author}{Rodr{\'\i}guez-Fortu{\~n}o,
  F.}, \bibinfo{author}{Nori, F.} \& \bibinfo{author}{Zayats, A.~V.}
\newblock \bibinfo{title}{Spin-orbit interactions of light}.
\newblock \emph{\bibinfo{journal}{Nature Photonics}}
  \textbf{\bibinfo{volume}{9}}, \bibinfo{pages}{796--808}
  (\bibinfo{year}{2015}).

\bibitem{liu2012compact}
\bibinfo{author}{Liu, Y.} \emph{et~al.}
\newblock \bibinfo{title}{Compact magnetic antennas for directional excitation
  of surface plasmons}.
\newblock \emph{\bibinfo{journal}{Nano Letters}} \textbf{\bibinfo{volume}{12}},
  \bibinfo{pages}{4853--4858} (\bibinfo{year}{2012}).

\bibitem{Pors2014}
\bibinfo{author}{Pors, A.}, \bibinfo{author}{Nielsen, M.~G.},
  \bibinfo{author}{Bernardin, T.}, \bibinfo{author}{Weeber, J.-C.} \&
  \bibinfo{author}{Bozhevolnyi, S.~I.}
\newblock \bibinfo{title}{{Efficient unidirectional polarization-controlled
  excitation of surface plasmon polaritons}}.
\newblock \emph{\bibinfo{journal}{Light Sci. Appl.}}
  \textbf{\bibinfo{volume}{3}}, \bibinfo{pages}{e197} (\bibinfo{year}{2014}).

\bibitem{yao2016controlling}
\bibinfo{author}{Yao, K.} \& \bibinfo{author}{Liu, Y.}
\newblock \bibinfo{title}{Controlling electric and magnetic resonances for
  ultracompact nanoantennas with tunable directionality}.
\newblock \emph{\bibinfo{journal}{ACS Photonics}} \textbf{\bibinfo{volume}{3}},
  \bibinfo{pages}{953--963} (\bibinfo{year}{2016}).

\bibitem{Kuznetsov2012}
\bibinfo{author}{Kuznetsov, A.~I.}, \bibinfo{author}{Miroshnichenko, A.~E.},
  \bibinfo{author}{Fu, Y.~H.}, \bibinfo{author}{Zhang, J.} \&
  \bibinfo{author}{Luk'yanchuk, B.}
\newblock \bibinfo{title}{{Magnetic light}}.
\newblock \emph{\bibinfo{journal}{Sci. Rep.}} \textbf{\bibinfo{volume}{2}},
  \bibinfo{pages}{492} (\bibinfo{year}{2012}).

\bibitem{evlyukhin2012demonstration}
\bibinfo{author}{Evlyukhin, A.~B.} \emph{et~al.}
\newblock \bibinfo{title}{Demonstration of magnetic dipole resonances of
  dielectric nanospheres in the visible region}.
\newblock \emph{\bibinfo{journal}{Nano Letters}} \textbf{\bibinfo{volume}{12}},
  \bibinfo{pages}{3749--3755} (\bibinfo{year}{2012}).

\bibitem{kuznetsov2016science}
\bibinfo{author}{Kuznetsov, A.~I.}, \bibinfo{author}{Miroshnichenko, A.~E.},
  \bibinfo{author}{Brongersma, M.~L.}, \bibinfo{author}{Kivshar, Y.~S.} \&
  \bibinfo{author}{Luk'yanchuk, B.}
\newblock \bibinfo{title}{Optically resonant dielectric nanostructures}.
\newblock \emph{\bibinfo{journal}{Science}} \textbf{\bibinfo{volume}{354}},
  \bibinfo{pages}{aag2472} (\bibinfo{year}{2016}).

\bibitem{fenollosa2008}
\bibinfo{author}{Fenollosa, R.}, \bibinfo{author}{Meseguer, F.} \&
  \bibinfo{author}{Tymczenko, M.}
\newblock \bibinfo{title}{Silicon colloids: from microcavities to photonic
  sponges}.
\newblock \emph{\bibinfo{journal}{Advanced Materials}}
  \textbf{\bibinfo{volume}{20}}, \bibinfo{pages}{95--98}
  (\bibinfo{year}{2008}).

\bibitem{XifrePerez2011}
\bibinfo{author}{Xifr\'{e}-P\'{e}rez, E.}, \bibinfo{author}{Fenollosa, R.} \&
  \bibinfo{author}{Meseguer, F.}
\newblock \bibinfo{title}{Low order modes in microcavities based on silicon
  colloids}.
\newblock \emph{\bibinfo{journal}{Opt. Express}} \textbf{\bibinfo{volume}{19}},
  \bibinfo{pages}{3455--3463} (\bibinfo{year}{2011}).
  
\bibitem{geffrin2012ncomm}
\bibinfo{author}{Geffrin, J.M.} \emph{et~al.}
\newblock \bibinfo{title}{Magnetic and electric coherence in forward- and back-scattered electromagnetic waves by a single dielectric subwavelength sphere}.
\newblock \emph{\bibinfo{journal}{Nature Communications}}
\textbf{\bibinfo{volume}{3}} (\bibinfo{year}{2012}).

\bibitem{fu2013ncomm}
\bibinfo{author}{Fu, Y.H.}, \bibinfo{author}{Kuznetsov, A.I.}, \bibinfo{author}{Miroshnichenko, A.E.}, \bibinfo{author}{Luk'yanchuk, B.S.}
\newblock \bibinfo{title}{Directional visible light scattering by silicon nanoparticles}.
\newblock \emph{\bibinfo{journal}{Nature Communications}}
\textbf{\bibinfo{volume}{4}} (\bibinfo{year}{2013}).

\bibitem{person2013demonstration}
\bibinfo{author}{Person, S.}, \emph{et~al.}
\newblock \bibinfo{title}{Demonstration of Zero Optical Backscattering from Single Nanoparticles}.
\newblock \emph{\bibinfo{journal}{Nano Letters}}
\textbf{\bibinfo{volume}{13}}, \bibinfo{pages}{1806--1809} (\bibinfo{year}{2013}).

\bibitem{yu2015high}
\bibinfo{author}{Yu, Y.~F.} \emph{et~al.}
\newblock \bibinfo{title}{High-transmission dielectric metasurface with 2$\pi$
  phase control at visible wavelengths}.
\newblock \emph{\bibinfo{journal}{Laser \& Photonics Reviews}}
  \textbf{\bibinfo{volume}{9}}, \bibinfo{pages}{412--418}
  (\bibinfo{year}{2015}).

\bibitem{Decker2015}
\bibinfo{author}{Decker, M.} \emph{et~al.}
\newblock \bibinfo{title}{High-efficiency dielectric huygens' surfaces}.
\newblock \emph{\bibinfo{journal}{Adv. Opt. Mater.}}
  \textbf{\bibinfo{volume}{3}}, \bibinfo{pages}{813--820}
  (\bibinfo{year}{2015}).

\bibitem{paniagua2016generalized}
\bibinfo{author}{Paniagua-Dom{\'\i}nguez, R.} \emph{et~al.}
\newblock \bibinfo{title}{Generalized brewster effect in dielectric
  metasurfaces}.
\newblock \emph{\bibinfo{journal}{Nature Communications}}
  \textbf{\bibinfo{volume}{7}} (\bibinfo{year}{2016}).

\bibitem{Evlyukhin2006JETP}
\bibinfo{author}{Evlyukhin, A.~B.} \& \bibinfo{author}{Bozhevolnyi, S.~I.}
\newblock \bibinfo{title}{Scattering of surface plasmon polaritons by a
  nanoparticle with the inclusion of the magnetic dipole contribution}.
\newblock \emph{\bibinfo{journal}{JETP Letters}} \textbf{\bibinfo{volume}{83}},
  \bibinfo{pages}{558--562} (\bibinfo{year}{2006}).

\bibitem{evlyukhin2015resonant}
\bibinfo{author}{Evlyukhin, A.~B.} \& \bibinfo{author}{Bozhevolnyi, S.~I.}
\newblock \bibinfo{title}{Resonant unidirectional and elastic scattering of
  surface plasmon polaritons by high refractive index dielectric
  nanoparticles}.
\newblock \emph{\bibinfo{journal}{Physical Review B}}
  \textbf{\bibinfo{volume}{92}}, \bibinfo{pages}{245419}
  (\bibinfo{year}{2015}).

\bibitem{li2017all}
\bibinfo{author}{Li, S.~V.} \emph{et~al.}
\newblock \bibinfo{title}{All-optical switching and unidirectional plasmon
  launching with electron-hole plasma driven silicon nanoantennas}.
\newblock \emph{\bibinfo{journal}{arXiv preprint arXiv:1703.03159}}
  (\bibinfo{year}{2017}).

\bibitem{tanemura2011multiple}
\bibinfo{author}{Tanemura, T.} \emph{et~al.}
\newblock \bibinfo{title}{Multiple-wavelength focusing of surface plasmons with
  a nonperiodic nanoslit coupler}.
\newblock \emph{\bibinfo{journal}{Nano Letters}} \textbf{\bibinfo{volume}{11}},
  \bibinfo{pages}{2693--2698} (\bibinfo{year}{2011}).

\bibitem{liu2011submicron}
\bibinfo{author}{Liu, J.~S.}, \bibinfo{author}{Pala, R.~A.},
  \bibinfo{author}{Afshinmanesh, F.}, \bibinfo{author}{Cai, W.} \&
  \bibinfo{author}{Brongersma, M.~L.}
\newblock \bibinfo{title}{A submicron plasmonic dichroic splitter}.
\newblock \emph{\bibinfo{journal}{Nature Communications}}
  \textbf{\bibinfo{volume}{2}}, \bibinfo{pages}{525} (\bibinfo{year}{2011}).

\bibitem{lee2014switchable}
\bibinfo{author}{Lee, S.-Y.}, \bibinfo{author}{Yun, H.}, \bibinfo{author}{Lee,
  Y.} \& \bibinfo{author}{Lee, B.}
\newblock \bibinfo{title}{Switchable surface plasmon dichroic splitter
  modulated by optical polarization}.
\newblock \emph{\bibinfo{journal}{Laser \& Photonics Reviews}}
  \textbf{\bibinfo{volume}{8}}, \bibinfo{pages}{777--784}
  (\bibinfo{year}{2014}).

\bibitem{lu2016integrated}
\bibinfo{author}{Lu, C.}, \bibinfo{author}{Liu, Y.-C.}, \bibinfo{author}{Hu,
  X.}, \bibinfo{author}{Yang, H.} \& \bibinfo{author}{Gong, Q.}
\newblock \bibinfo{title}{Integrated ultracompact and broadband wavelength
  demultiplexer based on multi-component nano-cavities}.
\newblock \emph{\bibinfo{journal}{Scientific Reports}}
  \textbf{\bibinfo{volume}{6}} (\bibinfo{year}{2016}).

\bibitem{Miroshnichenko2015Bianisotropy}
\bibinfo{author}{Miroshnichenko, A.~E.}, \bibinfo{author}{Evlyukhin, A.~B.},
  \bibinfo{author}{Kivshar, Y.~S.} \& \bibinfo{author}{Chichkov, B.~N.}
\newblock \bibinfo{title}{Substrate-induced resonant magnetoelectric effects
  for dielectric nanoparticles}.
\newblock \emph{\bibinfo{journal}{ACS Photonics}} \textbf{\bibinfo{volume}{2}},
  \bibinfo{pages}{1423--1428} (\bibinfo{year}{2015}).

\bibitem{sinev2016polarization}
\bibinfo{author}{Sinev, I.} \emph{et~al.}
\newblock \bibinfo{title}{Polarization control over electric and magnetic
  dipole resonances of dielectric nanoparticles on metallic films}.
\newblock \emph{\bibinfo{journal}{Laser \& Photonics Reviews}}
  \textbf{\bibinfo{volume}{10}}, \bibinfo{pages}{799--806}
  (\bibinfo{year}{2016}).

\bibitem{kerker1983electromagnetic}
\bibinfo{author}{Kerker, M.}, \bibinfo{author}{Wang, D.-S.} \&
  \bibinfo{author}{Giles, C.}
\newblock \bibinfo{title}{Electromagnetic scattering by magnetic spheres}.
\newblock \emph{\bibinfo{journal}{JOSA}} \textbf{\bibinfo{volume}{73}},
  \bibinfo{pages}{765--767} (\bibinfo{year}{1983}).

\bibitem{garcia2011directionality}
\bibinfo{author}{Garc{\'\i}a-C{\'a}mara, B.}, \bibinfo{author}{de~La~Osa,
  R.~A.}, \bibinfo{author}{Saiz, J.}, \bibinfo{author}{Gonz{\'a}lez, F.} \&
  \bibinfo{author}{Moreno, F.}
\newblock \bibinfo{title}{Directionality in scattering by nanoparticles:
  Kerker's null-scattering conditions revisited}.
\newblock \emph{\bibinfo{journal}{Optics Letters}}
  \textbf{\bibinfo{volume}{36}}, \bibinfo{pages}{728--730}
  (\bibinfo{year}{2011}).

\bibitem{hecht1996local}
\bibinfo{author}{Hecht, B.}, \bibinfo{author}{Bielefeldt, H.},
  \bibinfo{author}{Novotny, L.}, \bibinfo{author}{Inouye, Y.} \&
  \bibinfo{author}{Pohl, D.}
\newblock \bibinfo{title}{Local excitation, scattering, and interference of
  surface plasmons}.
\newblock \emph{\bibinfo{journal}{Physical Review Letters}}
  \textbf{\bibinfo{volume}{77}}, \bibinfo{pages}{1889} (\bibinfo{year}{1996}).

\bibitem{drezet2008leakage}
\bibinfo{author}{Drezet, A.} \emph{et~al.}
\newblock \bibinfo{title}{Leakage radiation microscopy of surface plasmon
  polaritons}.
\newblock \emph{\bibinfo{journal}{Materials Science and Engineering: B}}
  \textbf{\bibinfo{volume}{149}}, \bibinfo{pages}{220--229}
  (\bibinfo{year}{2008}).

\bibitem{zhang2010direct}
\bibinfo{author}{Zhang, D.~G.}, \bibinfo{author}{Yuan, X.} \&
  \bibinfo{author}{Bouhelier, A.}
\newblock \bibinfo{title}{Direct image of surface-plasmon-coupled emission by
  leakage radiation microscopy}.
\newblock \emph{\bibinfo{journal}{Applied Optics}}
  \textbf{\bibinfo{volume}{49}}, \bibinfo{pages}{875--879}
  (\bibinfo{year}{2010}).

\bibitem{frisbie2010characterization}
\bibinfo{author}{Frisbie, S.} \emph{et~al.}
\newblock \bibinfo{title}{Characterization of polarization states of surface
  plasmon polariton modes by fourier-plane leakage microscopy}.
\newblock \emph{\bibinfo{journal}{Optics Communications}}
  \textbf{\bibinfo{volume}{283}}, \bibinfo{pages}{5255--5260}
  (\bibinfo{year}{2010}).

\bibitem{zywietz2014laser}
\bibinfo{author}{Zywietz, U.}, \bibinfo{author}{Evlyukhin, A.~B.},
  \bibinfo{author}{Reinhardt, C.} \& \bibinfo{author}{Chichkov, B.~N.}
\newblock \bibinfo{title}{Laser printing of silicon nanoparticles with resonant
  optical electric and magnetic responses}.
\newblock \emph{\bibinfo{journal}{Nature Communications}}
  \textbf{\bibinfo{volume}{5}} (\bibinfo{year}{2014}).

\bibitem{dmitriev2016laser}
\bibinfo{author}{Dmitriev, P.} \emph{et~al.}
\newblock \bibinfo{title}{Laser fabrication of crystalline silicon
  nanoresonators from an amorphous film for low-loss all-dielectric
  nanophotonics}.
\newblock \emph{\bibinfo{journal}{Nanoscale}} \textbf{\bibinfo{volume}{8}},
  \bibinfo{pages}{5043--5048} (\bibinfo{year}{2016}).

\bibitem{Denisyuk2014}
\bibinfo{author}{Denisyuk, A.~I.}, \bibinfo{author}{Komissarenko, F.~E.} \&
  \bibinfo{author}{Mukhin, I.~S.}
\newblock \bibinfo{title}{{Electrostatic pick-and-place micro/nanomanipulation
  under the electron beam}}.
\newblock \emph{\bibinfo{journal}{Microelectron. Eng.}}
  \textbf{\bibinfo{volume}{121}}, \bibinfo{pages}{15--18}
  (\bibinfo{year}{2014}).

\bibitem{JohnsonChristy}
\bibinfo{author}{Johnson, P.~B.} \& \bibinfo{author}{Christy, R.~W.}
\newblock \bibinfo{title}{Optical constants of the noble metals}.
\newblock \emph{\bibinfo{journal}{Phys. Rev. B}} \textbf{\bibinfo{volume}{6}},
  \bibinfo{pages}{4370--4379} (\bibinfo{year}{1972}).

\end{thebibliography}

\newpage

\section*{Supplementary Information}

\subsection*{Details of the analytical model for excitation of SPP}

\subsubsection*{Derivation of SPP fields}

Here, we present the details of the derivation of equation for the fields of a surface plasmon polariton (SPP) excited by a nanoantenna with electric and magnetic dipole responses. We start from the equation for the magnetic field produced by point electric and magnetic dipoles (Eq.(1) in the main manuscript):

\begin{align}
\mathbf{H}(\mathbf{r})=k_0^2G_H(\mathbf{r})\mathbf{m}+ik_0\nabla\times G_E(\mathbf{r})\mathbf{p}.
\end{align}

Henceforth, without loss of generality, we restrict ourselves to the case of p-polarized excitation, when magnetic dipole along the $y$ axis, characterized by  magnetic polarizability $m_y$ and two electric dipoles along $z$ and $x$ axes with polarizabilities $p_z$ and $p_x$, are excited. The reflected parts of the Green functions can then be written explicitly as a two-dimensional Fourier transform:

\begin{align}
&{G}_H (\rho,\phi_0,0)=\frac{i}{8\pi^2}\iint k dk d\phi  \frac{r_p(k)}{k_z}\begin{pmatrix}\sin^2\phi & - \cos\phi\sin\phi & 0 \\ -\cos\phi\sin\phi & \cos^2\phi & 0 \\ 0 & 0 & 0 \end{pmatrix}e^{ik\rho\cos(\phi-\phi_0)+ik_zh} \label{eq:GHfull},\\&
\nabla \times {G}_E (r,\phi_0,0)=-\frac{1}{8\pi^2}\iint k dk d\phi  \frac{r_p(k)}{k_z}\begin{pmatrix}\sin 2\phi k_z/2 &  \sin^2\phi k_z & k\sin\phi \\ -\cos^2\phi k_z & -\sin 2\phi k_z/2  & -k \cos\phi \\ 0 & 0 & 0 \end{pmatrix}e^{ik\rho\cos(\phi-\phi_0)+ik_zh}\label{eq:GEfull},
\end{align}
where $k_z=\sqrt{k_0^2-k^2}$, and $r_p(k)=(k_z-k_{zm})/(k_z+k_{zm})$ is the magnetic field reflection coefficient for the TM-polarized plane wave with $k_{zm}=\sqrt{\varepsilon_m k_0^2-k^2}$, where $\varepsilon_m$ is the metal dielectric permittivity, and $h$ is the sphere radius. 

\begin{figure}
\includegraphics[width=\linewidth]{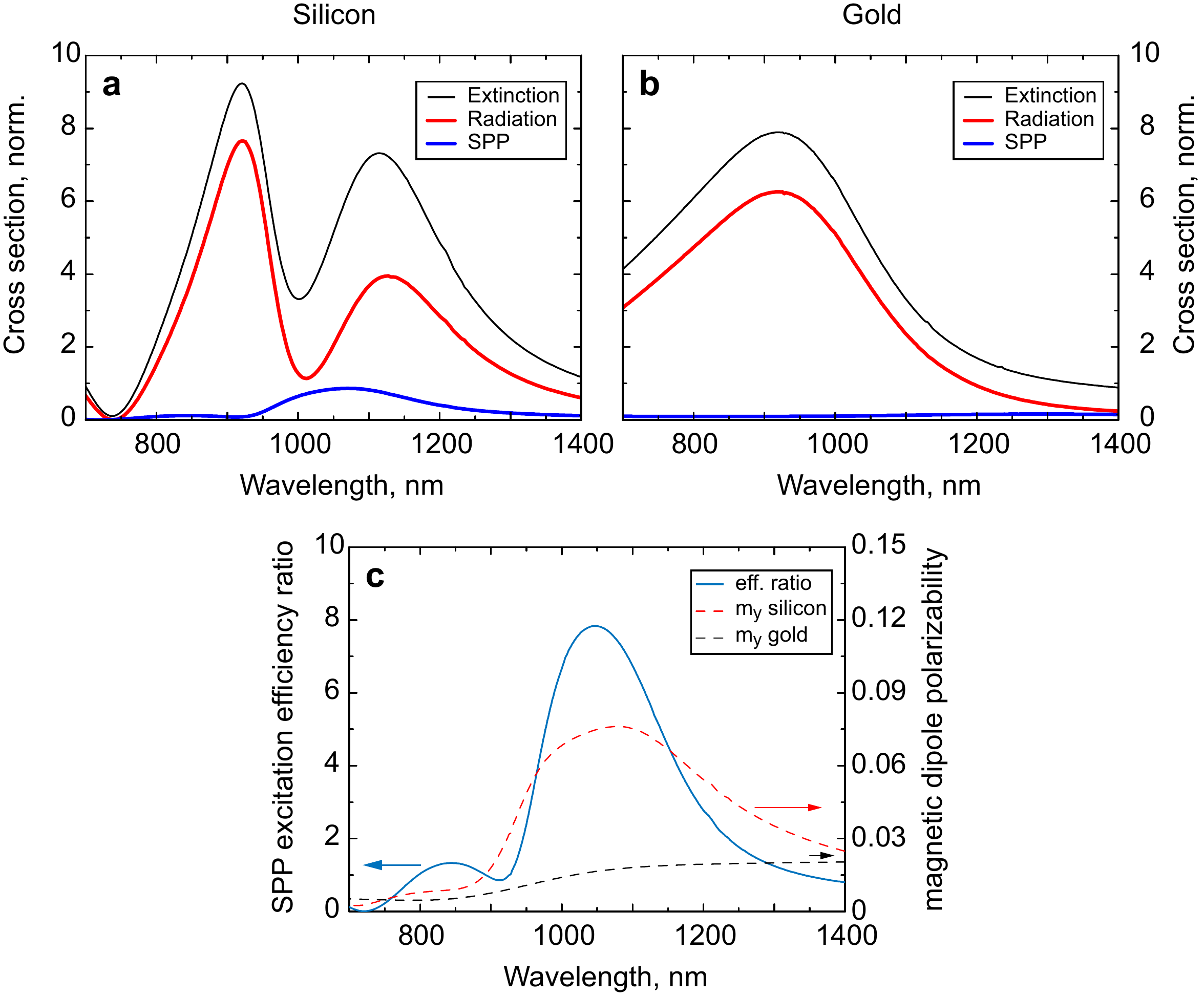}
\caption{\label{fig:CS} Top row: comparison of extinction (black line) and radiation (red line) cross sections with cross section of scattering into surface plasmon polariton (blue line) calculated analytically for a 297~nm silicon (a) and gold (b) nanospheres on gold substrate. The spheres are excited by a plane wave incident at 25~degrees to the substrate normal. All cross sections are normalized to the physical section of the nanosphere. Bottow row (c): spectral dependence of the ratio of SPP excitation with silicon and gold nanospheres (blue line) and their magnetic polarizabilities (dashed lines) for the same excitation geometry. The plots show superior SPP excitation efficiency with a silicon nanosphere close to its magnetic dipole resonance}
\end{figure}

The expressions \eqref{eq:GHfull},\eqref{eq:GEfull} can be reduced to the one-dimensional integrals in the straightforward manner by using the Jacobi-Anger expansion of the form $\int_0^{2\pi}\cos n\phi e^{ix\cos\phi}=2\pi(i)^nJ_n(x)$, where $J_n(x)$ is the Bessel function of the first kind. Then, using relation 

\begin{align}
J_n(x)=\frac{1}{2}\left(H_n^{(1)}(x)-(-1)^n H_n^{(1)}(-x)\right),
\end{align}

where $H_n^{(1)}(x)$ is the Hankel function of the first kind, the integration range can be transformed to the full real axis. Generally, the resulting integrals should be calculated numerically, however, since we are only interested in the surface plasmon field, which contributes as a residue of the pole of reflection coefficient $r_p(k)$ at $k=k_{SPP}=k_0\sqrt{\varepsilon_m/(\varepsilon_m+1)}$, we can obtain the analytical expression for the surface plasmon related contribution to the magnetic field:
\begin{align}
\mathbf{H}_{SPP}=\mathcal{N}\begin{pmatrix} \frac{1}{2}\sin 2\phi_0 H_2^{(1)}(k_{SPP}\rho)(m_y-\tilde{k}_zp_x) +i\tilde{k}_{SPP}\sin\phi_0 H_1^{(1)}(k_{SPP}\rho)p_z \\ \frac{1}{2}\left(H_0^{(1)}(k_{SPP}\rho)-\cos 2\phi_0 H_2^{(1)}(k_{SPP}\rho)\right)(m_y-\tilde{k}_zp_x)-i\tilde{k}_{SPP}\cos\phi_0 H_1^{(1)}(k_{SPP}\rho)p_z \\ 0\end{pmatrix},
\end{align}
where $\tilde{k}_{SPP}=k_{SPP}/k_0$, $\tilde{k}_z=\sqrt{1-\tilde{k}_{SPP}^2}$, $H_n^{(1)}$  are the $n$-th order Hankel functions of the first kind, and 
\begin{align}
\mathcal{N}=-\frac{1}{4}k_0^3\tilde{k}_ze^{ik_0\tilde{k}_zh}.
\end{align}
We then exploit the asymptotic form of Hankel function $H_n^{(1)}(z)\approx (\pi z/2)^{-1/2}\exp\left[i(z-\pi/4-n\pi/2)\right]$ to obtain
\begin{align}
\mathbf{H}_{SPP}\approx\mathcal{N}(\pi k_{SPP}\rho/2)^{-1/2}e^{ik_{SPP}\rho-i\pi/4}\begin{pmatrix} \frac{1}{2}\sin 2\phi_0(m_y-\tilde{k}_zp_x) +\tilde{k}_{SPP}\sin\phi_0p_z \\ \cos^2\phi_0(m_y-\tilde{k}_zp_x)-\tilde{k}_{SPP}\cos\phi_0p_z \\ 0\end{pmatrix}.\label{eq:HSPPSup}
\end{align}

The intensity of SPP in a given direction is proportional to $|H_{\phi}|^2$, where $H_{\phi}=H_y\cos\phi_0-H_x\sin\phi_0$. Thus, we arrive at the final expression for the SPP intensity (see Eq.(3) in the main manuscript):
\begin{align}
I_{SPP}\approx |\mathcal{N}|^2 \frac{2}{\pi k_{SPP} \rho} \left|\cos\phi_0 (m_y-i\kappa p_x) -\tilde{k}_{SPP}p_z\right|^2,
\end{align}

where $\kappa=-i\tilde{k}_z$. 

\subsubsection*{Calculation of SPP coupling efficiency}

To evaluate the efficiency of SPP excitation by a dielectric nanosphere, we first calculate the averaged plasmon-related Poynting vector:

\begin{align}
\langle S_{\rho}\rangle =\frac{1}{2}\mathrm{Re}\left(\mathbf{E}\times \mathbf{H}^*\right)\approx\frac{\mathrm{Re} ({\tilde{k}_{SPP}})}{2}|H_{\phi}|^2. \label{eq:SR}
\end{align}

Note that when obtaining Eq.\eqref{eq:SR} we omitted the propagation losses of a SPP by the substitution $k_{SPP}\rightarrow \mathrm{Re}\{k_{SPP}\}$ in the exponent argument in Eq.\eqref{eq:HSPPSup}.
We then integrate the Poynting vector over the upper half-space. We neglect the lower half space filled with metal since for the realistic cases when the surface plasmon is close to the light-line most of the energy of the surface plasmon is stored in the vacuum half-space. The SPP energy  $P_{SPP}$ is given after the integration by:
\begin{align}
\label{eq:PSPP}
P_{SPP}=\frac{k_0^4}{8}\mathrm{Re}({\kappa})e^{-2\mathrm{Re}(\kappa) k_0 h} \left(\frac{1}{2}|m_y-i\kappa p_x|^2+|\tilde{k}_{SPP}p_z|^2\right).
\end{align}
The SPP scattering cross-section is given by $\sigma_{SPP}=P_{SPP}/\langle S \rangle_{in}$, where $\langle S\rangle _{in}=|E_0|^2/2$.

The comparison of extinction and radiation cross sections of a 297~nm silicon nanosphere on gold substrate with cross section of scattering into the SPP calculated using Eq.\eqref{eq:PSPP} is presented in Fig.~\ref{fig:CS}a and reveals the strong contribution of the SPP to the optical response of the dielectric nanoparticles. At the same time, the contribution of the SPP to the extinction cross section of gold nanoparticles of the same size is very weak (see Fig.~\ref{fig:CS}b). The ratio of the SPP excitation efficiencies of these two nanoantennas reaches 8 close to the magnetic dipole resonace of the silicon particle (Fig.~\ref{fig:CS}c).

\subsection*{Setup for leakage radiation microscopy of SPP}

\begin{figure}
\includegraphics[width=0.4\linewidth]{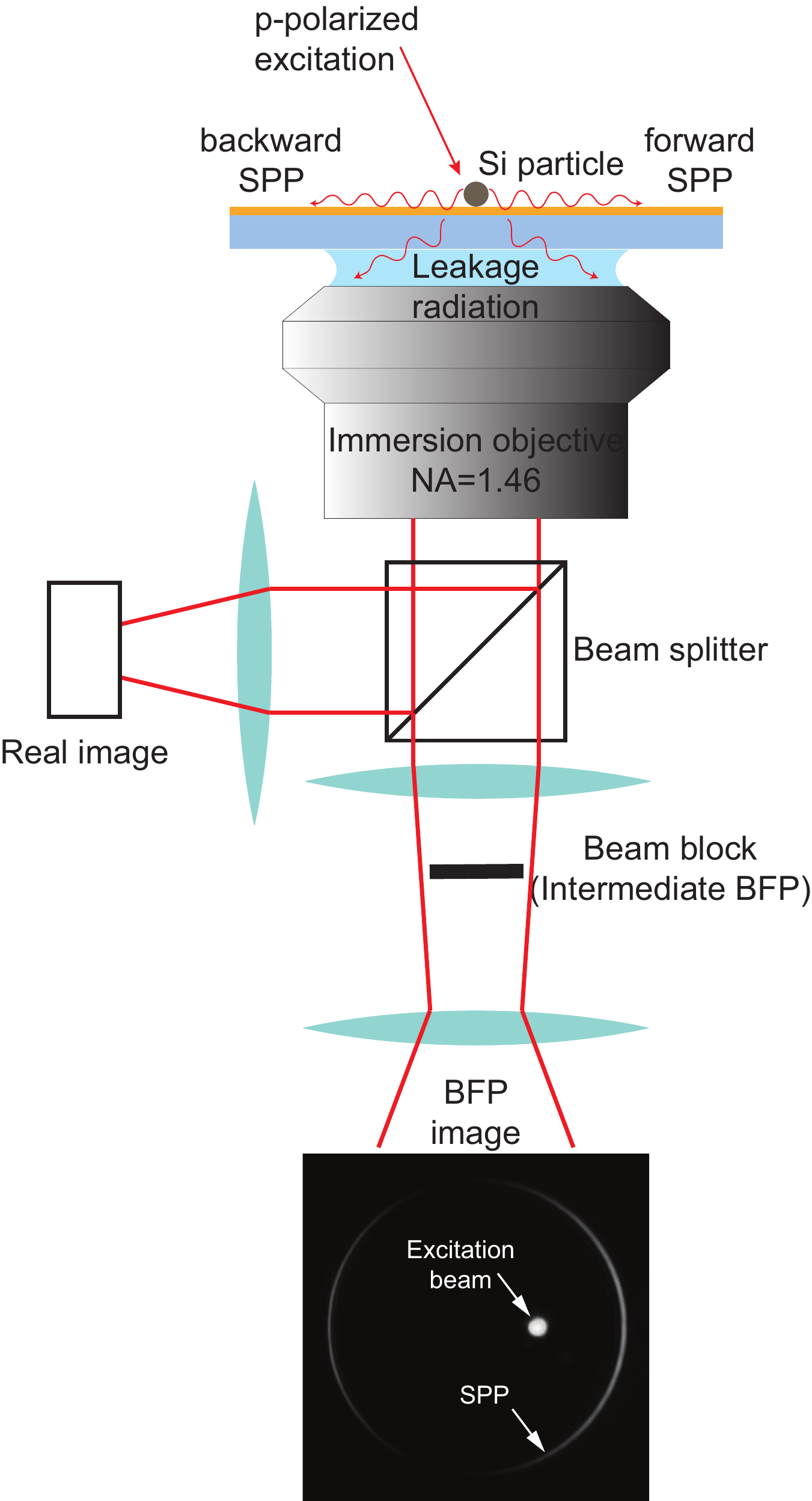}
\caption{\label{fig:expSetup} Scheme of the experimental setup for leakage radiation microscopy combined with Fourier plane imaging optics.}
\end{figure}

The scheme of the setup for measurements of directivity patterns of SPP excited by a single silicon nanosphere is presented in Fig.~\ref{fig:expSetup}. The SPP was excited with a TM-polarized beam incident at 25~degrees to the substrate normal and mildly focused with a 3~cm double achromat lens, which ensured low angular divergence of the excitation beam. The SPP radiation leaking through the thin gold film was collected from the bottom with an oil immersion objective (Zeiss 100x, NA=1.46). Both real and Fourier plane images of the sample were recorded simultaneously with two cameras after a non-polarizing beamsplitter. In the Fourier image channel, the incident beam was filtered with a 200~nm thick Cr disc placed in the intermediate Fourier plane of the optical system. An example of recorded Fourier image is shownin Fig.~\ref{fig:expSetup}. Leakage radiation of SPP excited by the sphere is responsible for the well-defined double-crescent pattern near the edges of the image.\cite{drezet2008leakage} At the same time, the excitation beam is still visible as a small spot to the left of the image center in Fig.~\ref{fig:expSetup}. This residual transmission  through the Cr disc allowed us to evaluate the angular broadening of the excitation beam, which amounted to approximately 6 degrees. However, in the Fourier images presented in the main manuscript the excitation beam was removed during the background subtraction process (reference images for subtraction were measured after moving the sphere out of the excitation spot).

\subsection*{Fourier representation of the SPP field}

In order to  simulate the experimental results obtained with  leakage radiation microscopy, we have modeled the SPP field distribution in the inverse Fourier space. The SPP mode can be characterized by its magnetic field component $H_{\varphi} (x,y)$ in the plane of metal-air interface. The directivity pattern shown in Fig. 4a,d in the main manuscript is obtained by  plotting the squared amplitude of Fourier component $|\hat{H}_{\varphi}(k_x,k_y)|^2$ given by the expression:
$$
\hat{H}_{\varphi}(k_x,k_y)=\int\int_S H_{\varphi}(x,y)\exp(ik_xx+ik_yy) dx dy.
$$
Here  the integration is taken over a circular area  $S$ of 15 $\mu$m radius in the plane  of metal-air interface. 

\subsection*{Captions for Movies S1-S3}

\subsubsection*{Movie S1}
Spectral dependence of the measured Fourier images of leakage radiation of SPP launched on 40~nm gold film by a $\approx$295~nm silicon nanosphere demonstrating the switching of SPP directivity. The sphere is excited by a mildly focused beam incident at 25~degrees to the surface normal. Top and bottom halves of the image corresponds to the excitation of SPP in the forward and backward directions, respectively.

\subsubsection*{Movie S2}
Calculated temporal dependence of the magnetic field near a 290~nm silicon nanosphere on 40~nm gold film on glass substrate excited by a plane wave (plane of incidence section is presented). The excitation wavelength and geometry ($\lambda$=935~nm, 25~degrees angle of incidence) correspond to backward (in the negative direction of the x~axis) SPP excitation.
Silicon nanosphere and gold layer are marked with white lines in the movie frames.

\subsubsection*{Movie S3}
Calculated temporal dependence of the magnetic field near a 290~nm silicon nanosphere on 40~nm gold film on glass substrate excited by a plane wave incident at 25~degrees to the substrate surface (plane of incidence section is presented). The excitation wavelength ($\lambda$=890~nm) correspond to forward (along the x~axis) SPP excitation. Silicon nanosphere and gold layer are marked with white lines on the movie frames.

\end{document}